\documentclass{nature}
\usepackage{graphicx}
\usepackage{mathrsfs}
\usepackage{amssymb}
\usepackage{amsmath}
\usepackage{textcomp}

\usepackage{graphicx,xcolor}
\usepackage{amsmath, bbm}
\usepackage{epstopdf}
\usepackage{color}

\usepackage{booktabs, bm}
\usepackage[super]{natbib}

\begin{document}
\bibliographystyle{naturemag.bst}

 \title{Strong plasmonic enhancement of biexciton emission:\\ controlled coupling of a single quantum dot to a gold nanocone antenna}
 \author{Korenobu Matsuzaki$^{*, 1}$, Simon Vassant$^{*, 1, 2}$, Hsuan-Wei Liu$^{1}$, Anke Dutschke$^{1,3}$, Bj\"orn Hoffmann$^{1}$, Xuewen Chen$^{1,4}$, Silke Christiansen$^{1,5}$, Matthew R. Buck$^{6}$, Jennifer A. Hollingsworth$^{6}$, Stephan G\"otzinger$^{1, 7}$ and Vahid Sandoghdar$^{1}$}
\maketitle
\begin{affiliations}
 \item Max Planck Institute for the Science of Light, Staudtstr. 2, D-91058 Erlangen, Germany. \item Present Address: SPEC, CEA, CNRS, Universit\'e Paris-Saclay, CEA/Saclay, 91191 Gif sur Yvette Cedex, France. \item Carl Zeiss Microscopy GmbH, Carl-Zeiss-Str. 22, 73447 Oberkochen, Germany. \item Present Address: School of Physics, Huazhong University of Science and Technology, Wuhan, People's Republic of China. \item Helmholtz-Zentrum Berlin f\"ur Materialien und Energie GmbH, Berlin, Germany.  \item Materials Physics \& Applications: Center for Integrated Nanotechnologies, Los Alamos National Laboratory, Los Alamos,
New Mexico 87545, USA \item Department of Physics and Graduate School of Advanced Optical Technologies, Friedrich Alexander University Erlangen-Nuremberg, D-91058 Erlangen, Germany
\newline Correspondence and requests for materials should be addressed to V.S. (email: vahid.sandoghdar@mpl.mpg.de).
\end{affiliations}
\newpage

\maketitle

\section{abstract}
Multiexcitonic transitions and emission of several photons per excitation comprise a very attractive feature of semiconductor quantum dots for optoelectronics applications. However, these higher-order radiative processes are usually quenched in colloidal quantum dots by Auger and other nonradiative decay channels. To increase the multiexcitonic quantum efficiency, several groups have explored plasmonic enhancement, so far with moderate results. By controlled positioning of individual quantum dots in the near field of gold nanocone antennas, we enhance the radiative decay rates of monoexcitons and biexcitons by 109 and 100 folds at quantum efficiencies of 60\% and 70\%, respectively, in very good agreement with the outcome of numerical calculations. We discuss the implications of our work for future fundamental and applied research in nano-optics.


\section{Introduction}
The ability to modify the lifetime of an atomic state by simply placing the atom in different environments continues to fascinate physicists \cite{Haroche-book:06}. Indeed, control of the optical properties of matter can also have exciting technological implications, e.g. in making brighter light emitting devices or more efficient lasers \cite{Yokoyama-book}. Since the first theoretical proposal of Purcell in 1946 \cite{purcell46}, a large body of experimental works has demonstrated that the spontaneous emission rate of an emitter can be modified close to surfaces \cite{Drexhage:74a}, in microresonators \cite{Berman-book, Yokoyama-book}, and in or close to nanostructures \cite{Pelton:15}. Nevertheless, very large enhancement factors of several hundreds or thousands remain a great challenge. In  particular, microcavity solutions, which require high quality factors, are not compatible with the broad spectra of solid-state emitters at room temperature. 

\begin{figure*}[b]
\centering 
 \includegraphics[width=.5\textwidth]{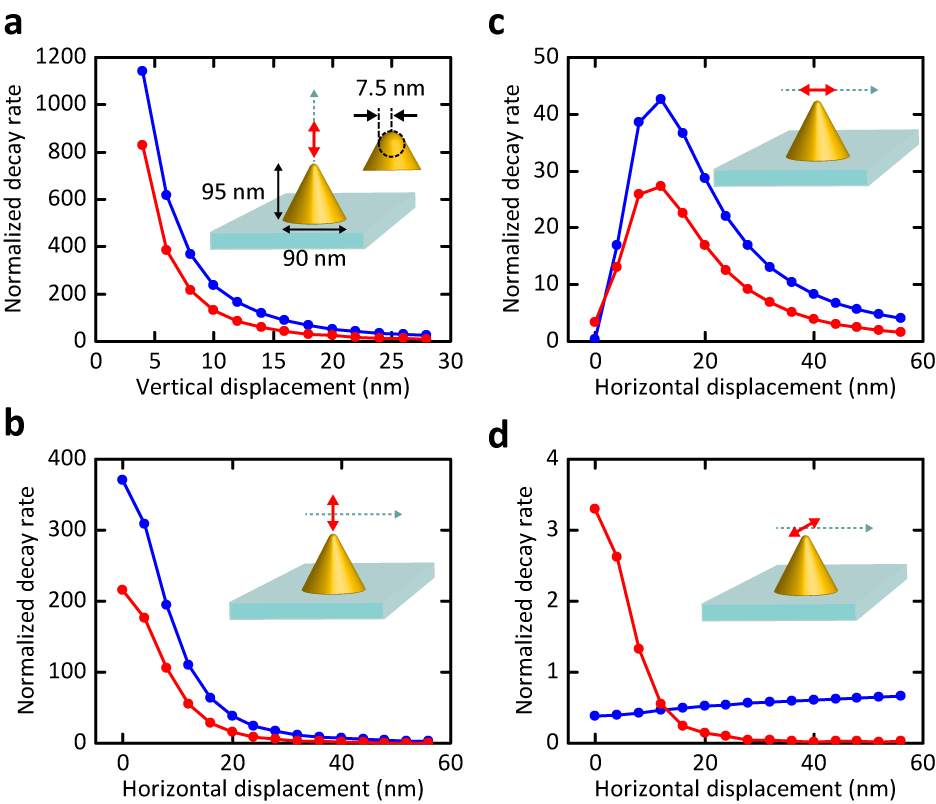}
 \caption{Radiative (blue) and nonradiative (red) decay rates of a point-dipole with different dipole moment orientations placed at different positions with respect to a gold nanocone fabricated on a glass substrate. The rates are normalized to that of an unperturbed dipole emitter in vacuum. The plasmon resonance was centered at about 625 nm and the transition wavelength was set at 650 nm to match the experimental data. a) Dipole moment and displacement along the cone axis. b) Varying lateral displacements for a dipole moment along the cone axis. c) Varying displacement in the direction of the dipole moment for a lateral dipole moment. d) Varying displacement normal to the direction of the dipole moment for a lateral dipole moment. The axial displacement of the qdot was 8\,nm for (b-d).} 
\label{theory}
\end{figure*}

The subwavelength dimension of plasmonic nanoantennas and their broad resonances offer an ideal solution for the modification of the emission properties of quantum emitters \cite{Pelton:15, Sandoghdar-antenna-book}. There are, however, technical difficulties. For example, fabrication of designer nanostructures with special form and size remains nontrivial. Moreover, both position and orientation of the emitter have to be adjusted with great precision to match the strongly inhomogeneous near fields of the plasmonic nanoantenna. A variety of methods such as chemical synthesis \cite{seelig07, Ji:15}, self-assembly \cite{Acuna:12}, lithographic nanofabrication \cite{curto10, Belacel:13, Meixner:15}, random distribution of emitters \cite{Song:05, kinkhabwala09, Punj:13, Zengin:13, Akselrod:14, Hoang:15} and use of scanning probe technology for nano-positioning \cite{Kuehn:06, anger06, Schietinger:09} have been employed over more than a decade to address these issues. Some works have reported plasmonic enhancement factors beyond one thousand times \cite{kinkhabwala09, Khatua:14}, but these refer to the overall fluorescence yield, combining the effects of excitation enhancement and improvement of quantum efficiency. Laboratory reports of very large spontaneous emission enhancements, on the other hand, are rare \cite{Lee:12, Eggleston:15, Hoang:16}, and their quantitative analyses leave room for improvement. 

Although fluorescence lifetime measurements are routine, distinction of the radiative ($\gamma_{\rm r}$) and nonradiative ($\gamma_{\rm nr}$) decay rates requires further information about the quantum efficiency defined as $\eta=\gamma_{\rm r}/(\gamma_{\rm r}+\gamma_{\rm nr})$ both in the absence and presence of the antenna. To assess $\eta$, one needs a precise knowledge of the excitation and emission rates, e.g., through careful saturation studies. Thus, such measurements demand a very high degree of photostability. Furthermore, because $\eta$ is very sensitive to the immediate environment of the emitter, it is imperative that one interrogates the very same emitter before and after coupling to an antenna if one hopes to extract a quantitative measure for the modifications of $\gamma_{\rm r}$ and $\gamma_{\rm nr}$.

Plasmonic enhancement brings about further complications when applied to semiconductor quantum dots (qdot) because it strongly affects the internal photophysics of the qdots \cite{Fernee:14}. For example, enhancement of the radiative rate can change the competition between the various decay channels such as Auger recombination and modify the blinking behavior \cite{Yuan:09, Ma:10, Canneson:12} and multiexcitonic emission dynamics \cite{LeBlanc:13, Park:13, Canneson:14, Wang:15}. Hence, a proper study of this rich landscape becomes particularly data intensive.

Our current work presents two important achievements. We report on more than one hundred fold enhancement of the spontaneous emission rate for a single qdot in the near field of a gold nanocone, while keeping a high quantum efficiency of 60\%. Moreover, we decipher monoexcitonic and biexcitonic emission processes of a single qdot and show a similarly high performance for the biexcitonic emission, corresponding to an improvement in the quantum efficiency of the latter by more than one order of magnitude. 

\section{Results}
\subsection{Theoretical predictions.} Figure \ref{theory} displays plots of the theoretical values of $\gamma_{\rm r}$ and $\gamma_{\rm nr}$ for an atom in vacuum interacting with a gold nanocone placed on a glass substrate normalized to $\gamma^{\rm vac}_{\rm r}$ in vacuum. In Fig.\,\ref{theory}a the emission dipole moment is parallel to the axis of the nanocone, and the axial displacement of the atom is varied. The data show that at a separation of 4 nm, $\gamma_{\rm r}$ and $\gamma_{\rm nr}$ are increased by about 1100 and 800, respectively, compared to $\gamma^{\rm vac}_{\rm r}$. This corresponds to a quantum efficiency of $\eta_{A} \simeq 60\%$ in the presence of the antenna if the quantum efficiency in its absence is assumed to be $\eta_0=1$. Here, the plasmon resonance was centered at about 625 nm, and the emitter transition wavelength was set at 650 nm to match the experimental parameters. We note that in the near infrared, nanocone antennas can result in spontaneous emission enhancements up to about $10000$ while keeping the quantum efficiency as high as 80\% \cite{mohammadi10, Chen:12}. 

In Fig. \ref{theory}b-d we also present $\gamma_{\rm r}$ and $\gamma_{\rm nr}$ for all three orthogonal orientations of the emission dipole as a function of the lateral separation from the cone tip. The data in Fig. 1 emphasize the sensitive dependence of the antenna effect on its relative position and orientation with respect to the emitter.  Ideally, one requires a single point-like quantum emitter with an emission dipole parallel to the nanocone antenna axis, which can be positioned in three dimensions with nanometer accuracy. Furthermore, it is important that the emitter be photostable to allow repeated measurements under different conditions.  

\begin{figure*}[t]
\centering 
 \includegraphics[width=0.95\textwidth]{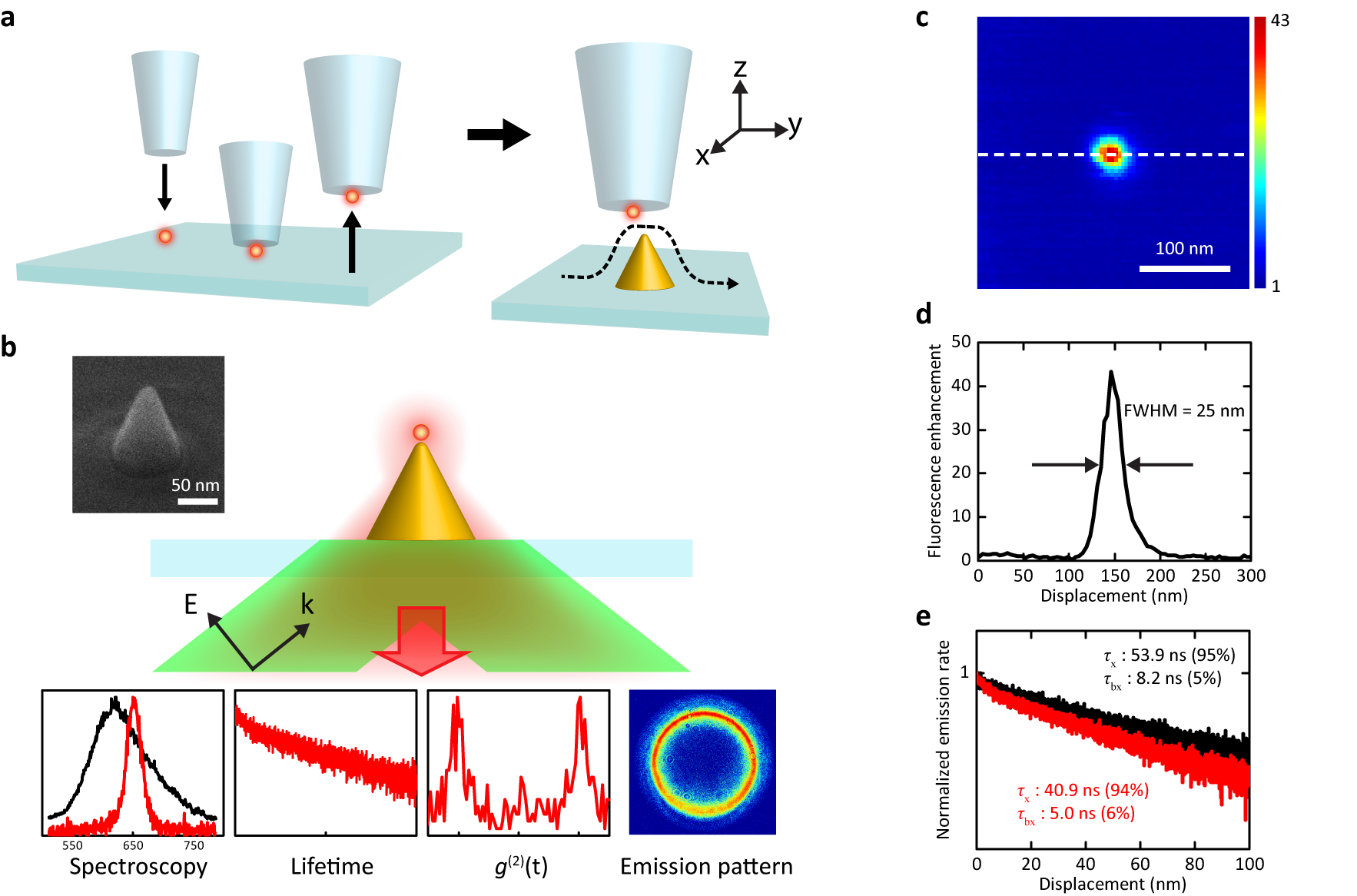}
 \caption{a) Left: Procedure for picking up a single quantum dot by a glass fiber tip. Right: Schematics of the tip scanning in the near field of a nanocone. b) Schematics of the total-internal reflection excitation and the detection of the red fluorescence analyzed in four different ways. The panels in the lower part show the spectra of the plasmon resonance (black, centered at 625 nm) and qdot emission (red) and sketch other modes of detection discussed in this paper. The inset shows a helium-ion microscope image of a nanocone. c) Map of the fluorescence signal recorded from the quantum dot as a function of its position with respect to the nanocone. d) A cross section from c). e) Fluorescence lifetime decay curves of a qdot on glass (black) and at the end of a glass fiber tip (red). The contributions of the monoexcitonic and biexcitonic emission lifetimes $\tau_{\rm x}$, $\tau_{\rm bx}$ and their weights (in parenthesis) were extracted from biexponential fits and are presented in the legend. }
\label{schematics}
\end{figure*}

\subsection{Experimental considerations.} In this work, we study the controlled coupling of a quantum dot and a cone nanoantenna placed on a glass substrate. As sketched in Fig. \ref{schematics}a, we used a shear-force microscope \cite{Karrai2000} to pick up and position individual qdots by a glass fiber tip \cite{Kalkbrenner:01}. Gold nanocones were fabricated with focused ion beam milling using Ga and He ions and characterized following the procedure reported in Ref.\,({\setcitestyle{numbers}\cite{Hoffmann:15}}). The inset in Fig \ref{schematics}b displays a helium-ion microscope image of such a cone. An oil-immersion microscope objective on the other side of the sample provided access to a wide range of optical measurements (see Fig. \ref{schematics}b). We used a picosecond pulsed laser at a wavelength of $\lambda=532$\,nm to excite the qdot in total internal reflection mode through the objective. To address the main antenna plasmon mode along the cone axis, we used p-polarized incident light. Distance-dependent studies were performed by positioning a selected nanocone under the qdot, which was kept fixed on the optical axis of the microscope objective. 

Figure \ref{schematics}c, d shows a lateral scan of the qdot fluorescence and a cross section from it, reaching an enhancement of about 45 within a full width at half-maximum (FWHM) of 25 nm. The fluorescence signal $S_0$ in the absence of the antenna is amplified according to the relation $S=K_{\rm exc}. K_{\eta}. K_{\xi}. S_0$ in the weak excitation limit, where $K_{\rm exc}$ stands for the enhancement of the excitation intensity at the position of the emitter, and $K_{\eta}$ and $K_{\xi}$ denote the antenna-induced modifications of the quantum efficiency and collection efficiency, respectively. Each of the $K$ values depends strongly on the dipolar orientation and position of the emitter with respect to the antenna, making it difficult to control and quantify at the single particle level. In addition, multiexcitonic emission has to be taken into account for qdots. 

A typical plasmon spectrum of our nanocones is shown in the lower left panel of Fig. \ref{schematics}b, showing that while it coincides well with the qdot emission spectrum, it is designed not to cause a substantial excitation enhancement at 532\,nm. Hence, we expect $K_{\rm exc}$ to be of the order of unity. In our experimental arrangement, $K_{\xi}$ also remains close to unity since we start with a high collection efficiency at a large numerical aperture (NA=1.4). We present a quantitative numerical analysis of this issue in the Methods section. Qdots in our current work were ``giant" quantum dots with a CdSe core and 16 shell layers of CdS  \cite{Chen:08}, which feature a nearly complete suppression of blinking and fluorescence intermittency \cite{Ghosh:12}. Ensemble measurements on these qdots indicate quantum efficiencies below or about 50\% {\cite{Vela:10} although in general $\eta_0$ can undergo significant variations at the single particle level \cite{Fernee:14, Orfield:16}. We now describe our procedure for determining $K_{\eta}$ and deciphering $\gamma_{\rm r}$ and $\gamma_{\rm nr}$ for both the monoexcitonic and biexcitonic emission pathways.

\begin{figure*}[t]
\centering 
 \includegraphics[width=0.9\columnwidth]{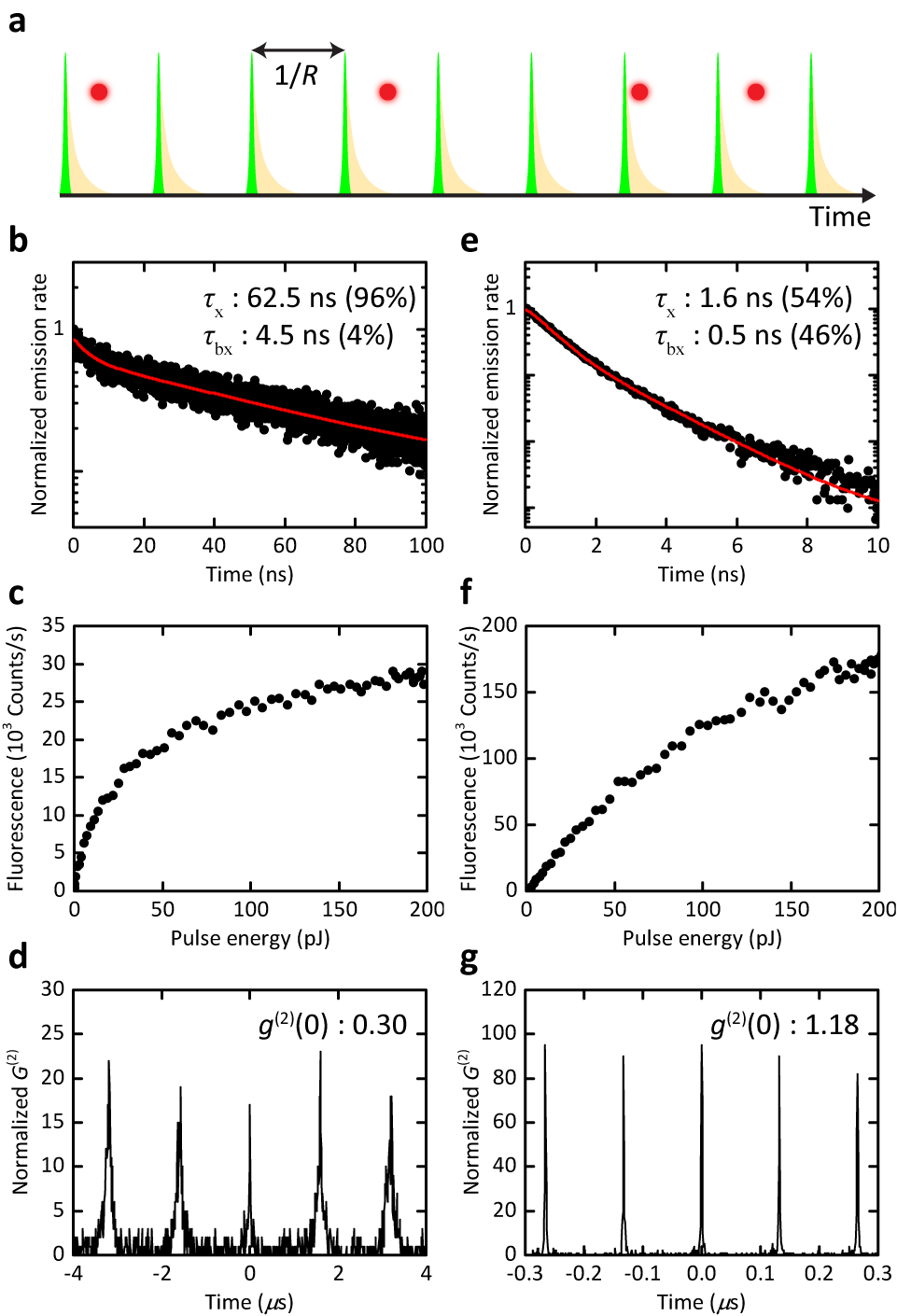}
\end{figure*}
\begin{figure*}[t]
 \caption{a) Schematic view of pulsed excitation. The red circles symbolize the sparsity of successful detection events. b) Fluorescence lifetime decay curve of a qdot at the end of a glass tip at low excitation power. c) Total fluorescence signal from the same qdot as a function of the excitation power. d) Autocorrelation function $g^{(2)}(0)$ of the same qdot at low excitation power. e) Fluorescence lifetime decay curve measured at low power for the same qdot as in (b) but in the near field of a gold nanocone. f) Total fluorescence signal from the same qdot on the nanocone as a function of the excitation power. g) Autocorrelation function $g^{(2)}(0)$ of the same qdot coupled to the nanocone at low excitation power. Excitation pulse repetition was 625 kHZ for (b, d), 3.75 MHz for (c, f), and 7.5 MHz for (e, g). Each curve was fit by two exponentials. The 1/e times of each component and their relative weights are displayed in each graph.}
\label{lifetime}
\end{figure*}

\subsection{Fluorescence lifetime: monoexciton and biexciton contributions.} To study the fluorescence lifetime decay, we excite the qdot by short laser pulses of 10\,ps and plot the number of detected photons as a function of delay after the pulse (see Fig. \ref{lifetime}a). Figure \ref{schematics}e displays an example of the fluorescence decay curve recorded from a qdot on a glass substrate (black) and after being attached to the fiber tip (red). The decay curves can be fitted by two fluorescence lifetimes $\tau$ ($1/e$ time) which we attribute to the monoexciton\,(x) emission path with a long lifetime and the biexciton\,(bx) channel with a short lifetime. The figure legend also shows the relative weights of each component according to the area under the two exponential curves used to fit the data. These measurements reveal that the transfer of the qdot from the substrate to the fiber tip modifies the decay rate. This observation emphasizes the sensitivity of qdots to their environment and conveys the important message that quantitative analyses of enhancement effects require comparison of the emission data from the very same qdot with and without the antenna.  

In Fig. \ref{lifetime}b, we display the measurement for another qdot attached to a glass fiber tip and approached to the glass substrate within the shear-force distance stabilization of a few nanometers. Here, we find monoexciton and biexciton components with 62\,ns and 4\,ns and weighting factors of 96\% and 4\%, respectively. Figure \ref{lifetime}e shows the fluorescence decay curve of the same qdot at the location of the highest fluorescence enhancement in the near field of a nanocone antenna. Again, the decay curve can be fitted with two exponential components, this time at $\tau=$1.6\,ns and 500\,ps with weighting factors of 54\% and 46\%, respectively. The measured fluorescence lifetime reports on the decay rate $\Gamma=1/\tau$ of the excited state population and is the sum of the radiative and nonradiative rates: $\Gamma=\gamma_{\rm r} + \gamma_{\rm nr}$. Thus, to decipher the components $\gamma_{\rm r}=1/\tau_{\rm r}$ and $\gamma_{\rm nr}=1/\tau_{\rm nr}$ separately, one needs to measure $\eta$. 

\begin{figure*}[t]
\centering 
 \includegraphics[width=0.9\textwidth]{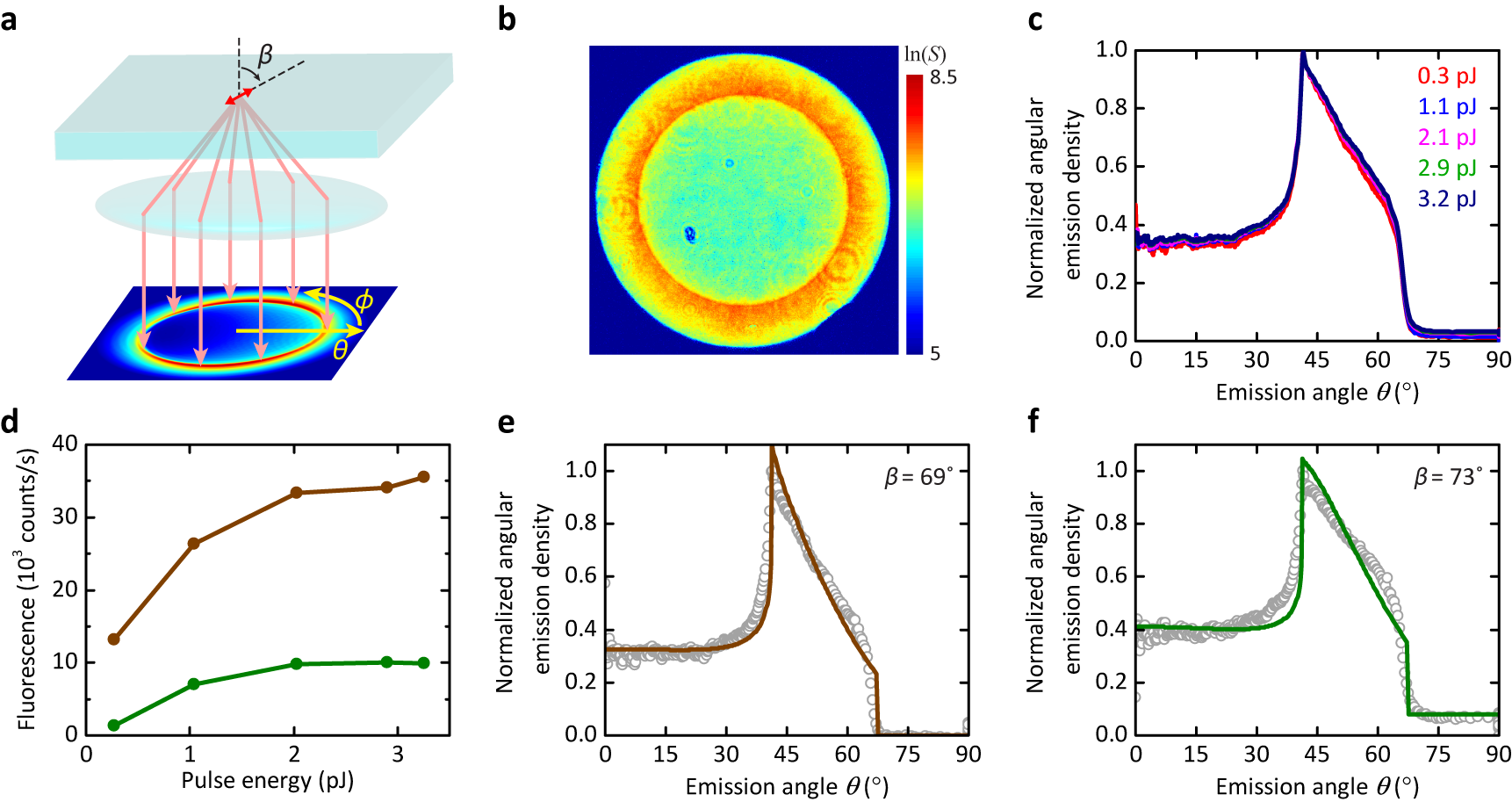}
 \caption{a) Schematics of back-focal plane measurements of the fluorescence distribution, which allows one to determine the angle $\beta$ of the dipole moment with respect to the optical axis of the setup. b) Back focal plane image of the emission from a single qdot placed on a glass substrate. c) Fluorescence at the back focal plane integrated over angle $\phi$ plotted as a function of angle $\theta$ for five different excitation powers. d) Monoexciton (brown) and biexciton (green) emission strengths extracted from fluorescence decay curves such as those in Fig. \ref{lifetime}. Note that the used incident powers are much smaller than those in Fig. \ref{lifetime}c,f because the laser beam was focused tightly in this case. e, f) The angular emission data deciphered for the exciton (e) and biexciton (f) contributions.}
\label{emission-pattern}
\end{figure*}

\subsection{Quantum efficiency: monoexciton and biexciton contributions.} The quantum efficiency $\eta$ is determined by the ratio of the number of emitted photons to the number of excitations. In our experiment, we drove the qdot in saturation to be sure that each incident laser pulse leads to an excitation event. Figure \ref{lifetime}c shows the total emission rate for the qdot attached to the fiber tip without the plasmonic nanocone as a function of the excitation power. This signal also includes the contribution of biexcitons, but by examining fluorescence decay curves recorded at different powers we confirmed that the contribution of the long-lifetime component did not change at powers beyond about 80 pJ/pulse. In other words, the monoexciton emission is saturated. The monoexciton fluorescence signal extracted from such analysis amounts to $S_0=3.5$ kcps at a laser repetition rate of $R=625$\,kHz. We, thus, deduce the number of emitted photons according to $S_0/\zeta$, where $\zeta=2.6\%$ stands for the overall detection efficiency of our setup. Next, we divided this quantity by $R$, whereby choosing very low $R$ values lets us ensure that each incident pulse finds the qdot in the ground state after the previous excitation, hence, eliminating complications posed by possible dark states. We, thus, arrive at $\eta^{\rm x}_{\rm 0}=22\%$ for the monoexciton quantum efficiency in the absence of the nanoantenna. Having determined $\eta^{\rm x}_{\rm 0}$, we can now extract the radiative and nonradiative lifetime $\tau^{\rm x}_{\rm 0, r}$ and $\tau^{\rm x}_{\rm 0, nr}$ of the unperturbed monoexciton to be 284\,ns and 80\,ns, respectively. 

To determine the biexciton quantum efficiency $\eta^{\rm bx}_{\rm 0}$, we resort to Hanbury-Brown and Twiss measurements, which allow us to record the second-order autocorrelation function $g^{(2)}(0)$ at zero time delay. By using the relation $g^{(2)}(0)=\eta^{\rm bx}/\eta^{\rm x}$, we can then extract the biexcitonic quantum efficiency \cite{Nair:11, Park:11}. The analysis of the areas under the pulses in Fig.\,\ref{lifetime}d lets us deduce $g^{(2)}_0(0)=0.3$, leading to $\eta^{\rm bx}_0=6.5\%$, $\tau^{\rm bx}_{\rm 0,r}$ = 69\,ns, and $\tau^{\rm bx}_{\rm 0,nr} $= 4.8\,ns on glass. 

\subsection{Enhancement factors: monoexciton and biexciton.} Next, we apply the same protocol to the data shown in Fig. \ref{lifetime}e-g for the qdot positioned on top of the cone. As shown in Fig. \ref{lifetime}f, the total fluorescence does not saturate within the available excitation power in our setup because of the enhanced contribution of higher order monoexcitons. Nevertheless, we can safely assume that the monoexciton population is again saturated at the maximum used power of 185 pJ/pulse because the nonresonant excitation rate in the absorption band of the qdot only depends on the incident power. Using the measured values of $\zeta=2.6\%$, $R=7.5$\,MHz, and monoexciton fluorescence signal $S=117.6$\,kcps, we find the monoexciton quantum efficiency $\eta^{\rm x}_A=60\%$ for the highest observed enhancement in the presence of the antenna. This analysis yields $\tau^{\rm x}_{\rm A, r}= 2.6$\,ns and $\tau^{\rm x}_{\rm A, nr}=4.1$\,ns, resulting in the radiative enhancement factor $\chi^{\rm x}_{\rm r}=\gamma^{\rm x}_{\rm A, r}/\gamma^{\rm x}_{\rm 0, r}=109$. Similarly, the $g^{(2)}(0)$ measurement in Fig. \ref{lifetime}g allows us to extract $\eta^{\rm bx}_A=71\%$, which in turn yields $\tau^{\rm bx}_{\rm A,r}$ = 0.69\,ns and $\tau^{\rm bx}_{\rm A,nr} $= 2.6\,ns when the qdot is coupled to the gold nanocone antenna. It follows that the radiative enhancement of the biexcitonic emission is $\chi^{\rm bx}_{\rm r}=100$. In a similar fashion, if we define $\chi_{\rm nr}=(\gamma_{\rm A, nr}-\gamma_{\rm 0, nr})/\gamma_{\rm 0, r}$ as a measure for the antenna-induced quenching rate, we find $\chi^{\rm x}_{\rm nr}=69$ and $\chi^{\rm bx}_{\rm nr}=26$. The results of this analysis are summarized in table \ref{table}.

\begin{table}
	\centering
	\normalfont
\scalebox{0.95}{
\begin{tabular}{c|cccccccc}
\toprule 
& $\gamma_{\rm 0,r}$ & $\gamma_{\rm 0,nr}$  & $\eta_{0}$ & $\gamma_{\rm A,r}$ & $\gamma_{\rm A,nr}$ & $\eta_{\rm A}$ & $\chi_{\rm r}$ & $\chi_{\rm nr}$\\
\hline 
x & $(284\, \rm ns)^{-1}$ & $(80\, \rm ns)^{-1}$ & 22\% & $(2.6\, \rm ns)^{-1}$ & $(3.9\, \rm ns)^{-1}$ & 60\% & 109 & 69\\
\hline 
bx & $(69\, \rm ns)^{-1}$ & $(4.8\, \rm ns)^{-1}$ & 6.5\% & $(0.69\, \rm ns)^{-1}$ & $(1.7\, \rm ns)^{-1}$ & 71\% & 100 & 26\\
\bottomrule
\end{tabular}}
\caption{Summary of the outcome of the analysis of the photophysics of a quantum dot before and after coupling to a plasmonic nanocone antenna.}
\label{table}
\end{table}

Our finding that the monoexciton and biexciton emission rates are enhanced by about the same factor is consistent with the assumption that both processes can be attributed to a dipolar transition. To examine this hypothesis, we studied the angular emission pattern of qdots placed on a glass substrate. Here, we recorded the fluorescence distribution of a single qdot in the back focal plane of the microscope objective as presented in Fig. \ref{emission-pattern}a, b. The red curve in Fig. \ref{emission-pattern}c plots the distribution as a function of angle $\theta$ after integrating the data over angle $\phi$. The other curves in Fig. \ref{emission-pattern}c show that the pattern does not change as the biexciton contribution increases at different excitation powers. To examine this hypothesis more closely, we recorded fluorescence decay curves at five different powers to separate the contributions of the monoexcitons and biexcitons as discusses above. The outcome is shown in Fig. \ref{emission-pattern}d. Assuming that the dipole moments of the monoexcitons and biexcitons do not change as a function of the excitation power, we fitted all five angular emission data sets simultaneously and arrived at the emission patterns displayed in Fig. \ref{emission-pattern}e, f for the monoexciton and biexciton channels, confirming nearly identical dipole orientations.

\subsection{Distance dependence of monoexciton and biexciton enhancement.} In this section, we report on position-dependent studies to visualize the evolution of the monoexciton and biexciton emission modification. Figure \ref{lateral-scan}a,b displays the long and short lifetime components of the fluorescence decay curves as a qdot was laterally displaced away from the cone apex. In Fig. \ref{lateral-scan}c we present the same data in the normalized fashion together with the values of $g^{(2)}(0)$ recorded at each point. The growth of the latter from less than 0.2 far from the antenna to slightly higher than 1 at the cone clearly indicates the transition from single-photon emission to the emission of two or more photons per excitation pulse. This behavior is also mirrored in Fig. \ref{lateral-scan}d, which plots the evolution of the relative weights of the two components of the biexponential fits to approximately equal amounts. 

We can use the data in Fig. \ref{lateral-scan}a-d to determine $\alpha= \gamma^{\rm bx}_{\rm 0, r}/\gamma^{\rm x}_{\rm 0, r}$ as an intrinsic property of a qdot.  Theory suggests that the value of $\alpha$ depends on the spin-flip rate with $\alpha=2$ for the case of slow spin flip and $\alpha=4$ for fast spin flips \cite{Narvaez:06, Klimov:08}. To determine $\alpha$, we note that  one can formulate the compact relation $(\Gamma^{\rm bx}_ {\rm A} - \Gamma^{\rm bx}_0)=\alpha (\Gamma^{\rm x}_{\rm A}- \Gamma^{\rm x}_0)$, which should hold at every qdot-nanocone distance (see the Methods section). As displayed in Fig. \ref{lateral-scan}e, the series of data in Fig. \ref{lateral-scan} confirms such a linear relationship with $\alpha=2.2$, while the data in table \ref{table} recorded on a different qdot yield $\alpha=4.1$. 
\begin{figure*}[t]
\centering 
 \includegraphics[width=\textwidth]{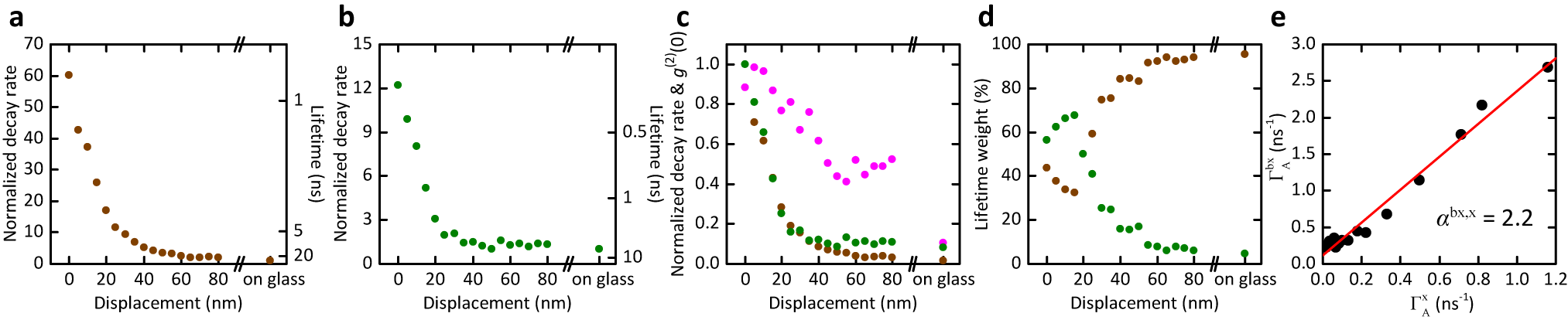}
 \caption{Lateral distance dependence of the monoexciton (a) and biexciton (b) fluorescence decay rates and lifetimes of a qdot. Distance zero denotes the cone apex. c) The autocorrelation function $g^{(2)}(0)$ measured at each lateral displacement (pink). The same data as in (a) and (b) are also plotted normalized to their maximum values. d) The distance dependence of the weighting ratios of the long and short lifetime components attributed to the monoexciton and biexciton emission channels. The measured values far from the nanocone and close to a glass substrate are also shown in each case. e) The measured values of $\gamma^{\rm bx}_{\rm 0,r}$ and $\gamma^{\rm x}_{\rm 0,r}$, leading to the slope $\alpha$. }
\label{lateral-scan}
\end{figure*}

\subsection{Towards a monolithic hybrid system.}
Although in our experiment we have focused on the controlled positioning of a single qdot to obtain quantitative data, new nanofabrication techniques can be used to place single qdots at the cone apex to construct composite devices \cite{Meixner:15, Kress:15}. In our laboratory, we realized such a hybrid structure in a preliminary fashion by mechanically transferring the qdot from the tip onto the cone. Figure \ref{dot-on-cone}a displays the emission pattern of the fluorescence recorded in the back-focal plane, representing an effective dipole moment with a small tilt angle of $\beta=0.5^\circ$ with respect to the cone axis. Interestingly, as shown in Fig. \ref{dot-on-cone}b, the emission pattern of the same qdot on the glass substrate was clearly not axial. In other words, the antenna resonance dominates the emission pattern of the qdot. We emphasize that this effect was reproducible on many qdots. Similar behavior has also been reported for rod antennas \cite{taminiau08a}. 

\section{Discussion and future prospects} 

Theoretical calculations indicate that radiative enhancement factors as large as several thousands are within reach with nanocone antennas if one tunes the wavelength of interest to the near infrared to minimize the losses in gold \cite{mohammadi10, Chen:12}. In our experiment, the design of the cones for the spectral domain of the used emitters (see Fig. \ref{theory}a) as well as several technical issues limit the experimentally obtained factors. In particular, the qdot radius of about 8\,nm restricts the separation between the emission dipole and the cone apex, implying a maximum value of $\chi_{\rm r}=350$ and $\eta=57\%$ for an axial dipole moment. Furthermore, as seen in Fig. \ref{emission-pattern}, qdot dipole moments are in general not oriented axially. For example, a tilt of $\beta=60^\circ$ would reduce the enhancement factor to 93. 

A hundred-fold enhancement of the biexcitonic emission at a quantum efficiency of about 70\% opens the doors to many applications in light emitting technologies \cite{Malko:02, Kazes:02, Dang:12, Shirasaki:13, Eggleston:15}. Indeed, the small size of the composite qdot-nanocone structure lends itself to integration in other structures such as microcavities or planar antennas \cite{Lee:11, Chu:14}, e.g. for achieving near-unity collection efficiency and brighter photon sources. In addition to the enhancement of incoherent fluorescence, large radiative enhancements are also very promising for fundamental solid-state spectroscopy and quantum optics because enhancement of $\gamma_{\rm r}$ directly translates to a larger extinction cross section given by $\sigma=\gamma_{\rm r}/(\gamma_{\rm r}+\gamma_{\rm nr}+ \gamma_{\rm deph})$, where $\gamma_{\rm deph}$ denotes the dephasing rate. Because at room temperature this quantity is larger than $\gamma_r$ by nearly five orders of magnitude, enhancement of $\gamma_{\rm r}$ by several thousand folds would directly translate to a similar enhancement of $\sigma$ \cite{Kukura:09}. Thus, large radiative enhancements will help usher in a new era of coherent plasmonics \cite{Chen:13, Chikkaraddy:16}. 

\begin{figure*}[t]
\centering 
 \includegraphics[width=\columnwidth]{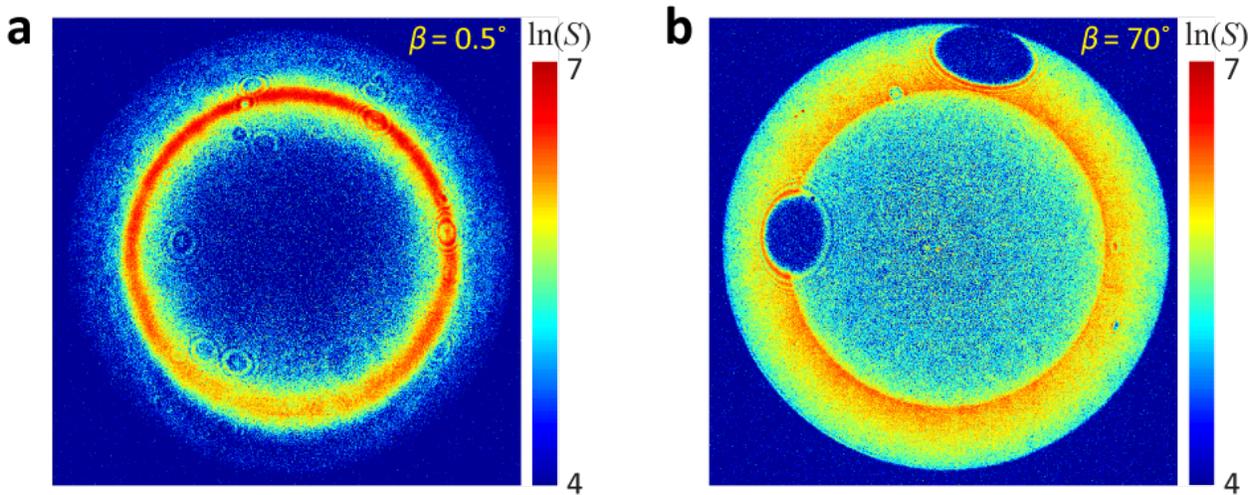}
 \caption{a) The back-focal plane image of the fluorescence of a qdot deposited at the apex of a nanocone antenna. b) Same as in (a) but for a qdot on glass before being picked up.}
\label{dot-on-cone}
\end{figure*}

\textbf{Acknowledgments}
We thank Maksim Schwab for his efforts in the mechanical workshop. This work was supported by an Alexander von Humboldt professorship, the Max Planck Society and the European Research Council (Advanced Grant SINGLEION). The giant quantum dot (qdot) samples were provided by J.A.H. through a Center for Integrated Nanotechnologies (CINT) User Project (C2015B0014), where CINT is a U.S. Department of Energy (DOE), Office of Basic Energy Sciences (OBES) Nanoscale Science Research Center and User Facility, and giant qdot development is funded through a U.S. DOE, OBES Division of Materials Science and Engineering grant (2009LANL1096). 

\textbf{Author contributions}
V.S. designed and conceived the experiments. K.M., H-W.L., and S.V. performed the measurements. A.D. and B.H. fabricated the nanocones. S.C. supervised the work done by B.H. Theoretical calculations were performed by X-W.C. and H-W.L. The quantum dot samples were provided by M.R.B and J.A.H. The experimental work was supervised by S.G. and V.S. The paper was written by V.S., K.M. and H-W. L. and was commented by S.G., X-W.C., A.D., B.H., and J.A.H. In this work, K.M. and S.V. contributed equally.
	

\section{Methods}

\subsection{Optical measurements}

As a light source, we used the frequency-doubled output of a passively mode-locked Nd: YVO$_4$ oscillator (Time-Bandwidth, Cheetah-X) with a repetition rate of 75 MHz, pulse duration of 10 ps, and output power of 2 W. In order to reduce the repetition rate, the laser output was sent through a pulse picker (APE, PulseSelect). Depending on the lifetime of the emitter, the repetition rate after pulse picking was adjusted to $R =$\ 625 kHz - 7.5 MHz. The picked pulses were introduced into a single mode fiber in order to obtain a good beam pattern, and were passed through a short pass filter (840 nm) and a laser line filter (532 nm) to completely isolate the emission line at 532 nm. 

The excitation light was sent to an oil-immersion objective lens (Olympus, UPlanSApo, 100x, NA 1.4) in total internal reflection geometry with P-polarization to ensure that we have a large electric field component in the vertical direction. The fluorescence emission of the sample was collected by the same objective lens and was separated from the excitation path using a 1:1 beam splitter. The residual excitation light was eliminated by the combination of a long pass filter (550 nm) and a bandpass filter (center = 655 nm, width = 40 nm, the wavelength was optimized for each emitter by adjusting the incident angle). 

For the back focal plane imaging, the fluorescence at the back focal plane of the objective lens was projected onto a scientific CMOS camera (Hamamatsu, ORCA-Flash 4.0 v2). 
In order to address a single emitter, the excitation beam was tightly focused onto the sample plane in this measurement. 

For the lifetime and $g^{(2)}$ measurements, a wide-field illumination was used to facilitate the measurements, while a variable pinhole was placed in the detection path so that only the light emitted from a single emitter was passed through. 
Using a 1:9 beam splitter, 10\% of the fluorescence was sent to an EMCCD camera (Hamamatsu, ImagEM Enhanced) for observing the fluorescence image. 
The remaining 90\% of the fluorescence was further split by a 1:1 beam splitter and was detected by APD1 (MPD, PD-050-CTB) and APD2 (ID Quantique, ID100-50). 
The two APDs were connected to a time-correlated single-photon counting (TCSPC) unit (PicoQuant, HydraHarp), which enabled us to construct fluorescence decay curves (using only APD1) and $g^{(2)}(t)$ curves (using both APD1 and APD2) simultaneously. 
The overall detection efficiency for APD1 was $\zeta =$\ 2.6\%, which was determined from the transmission through all the optics (10.5\%), the quantum efficiency of APD1 (37\% at 650 nm), and the collection efficiency of the objective lens ($\xi = 67\%$ based on the result of a simulation. See the description below for details.).

\subsection{Simulation}
\label{sec:simulation}
Three-dimensional numerical simulations were performed with finite-difference time-domain method (FDTD Solutions, Lumerical Solutions). We set the dimensions of the gold nanocone to a height of 95 nm, base diameter of 90 nm, and tip radius of 7.5 nm (see Fig. \ref{theory}a). It was placed on a glass substrate, and a radiating dipole with an unperturbed quantum efficiency of unity was positioned in the vicinity of the nanocone. The dielectric function of gold was modeled using the experimental data reported in the literature \cite{CRC-2006}, and the refractive index of the glass substrate was set to 1.5.
The system was surrounded by perfectly matched layer (PML) boundaries (2000 nm $\times$ 2000 nm $\times$ 2000 nm) centered at the position of the emission dipole. The distance between the boundary and the emitter was chosen to be large enough to avoid the absorption of near fields by the PML boundaries. The finest mesh size was set to 1 nm to achieve sufficient simulation accuracy within reasonable memory requirement. 

The decay rate enhancement of an emitter by the plasmonic nanocone antenna was evaluated by considering the power emitted by an oscillating point-like dipole in the presence of the nanocone and normalizing it with respect to the case in vacuum \cite{Agio:12}, satisfying the relationships
\begin{align}
	\frac{\gamma_{A, r}}{\gamma^{\rm vac}_{r}} = \frac{P_{A, r}}{P^{\rm vac}_{r}},\quad \frac{\Gamma_{A}}{\gamma^{\rm vac}_{r}} = \frac{P_{A, tot}}{P^{\rm vac}_{r}},
\end{align}
where $\gamma_{A, r}$, $\gamma^{\rm vac}_{r}$ denote the radiative decay rates in the presence of the nanocone antenna and in vacuum, $P_{A, r}$ and $P^{\rm vac}_{r}$ are the power radiated to the far-field in the presence of the nanocone antenna and in vacuum, $\Gamma_{A}$ is the total decay rate with the nanocone, and $P_{A, tot}$ is the total power dissipated by the dipole with the nanocone, i.e. including the part absorbed by the gold nanocone and that radiated. The radiative decay rate enhancement ($\gamma_{A, r}/\gamma^{\rm vac}_{r}$) 
at each position was then determined by the calculated values of $P_{A, r}$ and $P^{\rm vac}_{r}$ at that position. In order to study the losses caused by the nanocone, we also evaluated the normalized nonradiative decay rate ($\gamma_{A, nr}/\gamma^{\rm vac}_{r}$) at each position. 
This was done by subtracting the radiative decay rate from the total decay rate:
\begin{align}
	\frac{\gamma_{A, nr}}{\gamma^{\rm vac}_{r}} 
	= \frac{\Gamma_{A}}{\gamma^{\rm vac}_{r}} - \frac{\gamma_{A, r}}{\gamma^{\rm vac}_{r}}
	= \frac{P_{A,tot}}{P^{\rm vac}_{r}} - \frac{P_{A, r}}{P^{\rm vac}_{r}}.
\end{align}

For the simulation of the emission pattern, a tapered glass tip (refractive index is 1.5) with a plateau diameter of 150 nm and an opening angle of $30^\circ$ was placed 5 nm above the dipole which was positioned 8 nm above the apex of the nanocone or the surface of the glass substrate. The radiating near fields in the lower half-space were projected to the far field by far-field transformation. The resulting electric field distribution was then transformed into an angular emission pattern. 

\subsection{Collection efficiency}
\label{sec:collectionEfficiency}
The collection efficiency of an objective is defined as $\xi=P_{\rm coll}/P_{\rm r}$, where $\xi$, $P_{\rm coll}$, and $P_{\rm r}$ are the collection efficiency, the power collected by the objective lens and the total power radiated over the whole solid angle, respectively. Here, $P_{\rm coll}$ was evaluated using an emission pattern obtained from the numerical simulation both in the absence and in the presence of a nanocone antenna within the collection angle of 67.25$^\circ$, corresponding to the numerical aperture of 1.4. Combined with $P_{\rm r}$, which was also calculated with a numerical simulation, the collection efficiency with and without the nanocone was estimated.

In particular, for the typical dipole orientation of our emitter ($\theta_{\rm dipole}=60^\circ$), $\xi_{0}$ and $\xi_{A}$ were determined to be 67.7\% and 66.9\%, respectively. This result shows that a nanocone antenna does not cause a large modification of the collection efficiency ($K_\xi=\xi_{A}/\xi_{0}=$ 0.99). For completeness, because the emission dipole orientation of the emitter is not controlled in our experiment, we examined $\xi_0$, $\xi_A$, and $K_\xi$ with various dipole orientations ($\theta_{\rm dipole}=0^\circ$, 20$^\circ$, 40$^\circ$, 60$^\circ$, and 80$^\circ$).
The simulation yielded $66.3\% < \xi_0 < 69.7\%$, $66.7\% < \xi_A < 66.9\%$, and $0.96 < K_\xi < 1.01$, which shows that the collection efficiency is not very sensitive to the dipole orientation and $K_\xi$ is always close to unity in our system.

\subsection{Derivation of $\bm{(\Gamma^{\rm bx}_A - \Gamma^{\rm bx}_0) = \alpha (\Gamma^{\rm x}_A - \Gamma^{\rm x}_0)}$}
As discussed in the main text, the total decay rates with and without an antenna can be expressed respectively as 
\begin{align}
	\Gamma_0 &= \gamma_{0,r} + \gamma_{0,nr} \label{eq:gamma_0_tot},\\
	\Gamma_A &= \gamma_{A,r} + \gamma_{A,nr} \label{eq:gamma_A_tot},
\end{align}
with
\begin{align}
	\gamma_{A,r} &= \chi_{\rm r} \gamma_{0,r} \label{eq:gamma_A_r},\\
	\gamma_{A,nr} &= \gamma_{0,nr} +  \chi_{\rm nr} \gamma_{0,r} \label{eq:gamma_A_nr}. 
\end{align}
By combining these equations and eliminating $\gamma_{0,nr}$, the total decay rate in the presence of the antenna can be expressed as 
\begin{align}
	\Gamma_A 
	&= \chi_{\rm r} \gamma_{0,r} + \left\{ (\Gamma_0 - \gamma_{0,r}) + \chi_{\rm nr} \gamma_{0,r} \right\}\\
	&= \left\{\chi_{\rm r} + \chi_{\rm nr} - 1\right\} \gamma_{0,r} + \Gamma_0. 
\end{align}
This expression is valid for both monoexciton and biexciton emissions. 
\begin{align}
	\Gamma^{\rm x}_A - \Gamma^{\rm x}_0 &= \left\{\chi_{\rm r} + \chi_{\rm nr} - 1\right\} \gamma^{\rm x}_{0,r},\\
	\Gamma^{\rm bx}_A - \Gamma^{\rm bx}_0 &= \left\{\chi_{\rm r} + \chi_{\rm nr} - 1\right\} \gamma^{\rm bx}_{0,r}. 
\end{align}
Here, the factor $\left\{\chi_{\rm r} + \chi_{\rm nr} - 1\right\}$ is identical in the two equations. 
Thus, after the elimination of this common factor, we obtain 
\begin{align}
	\Gamma^{\rm bx}_A - \Gamma^{\rm bx}_0 &= \frac{\gamma^{\rm bx}_{0,r}}{\gamma^{\rm x}_{0,r}} (\Gamma^{\rm x}_A - \Gamma^{\rm x}_0)=\alpha(\Gamma^{\rm x}_A - \Gamma^{\rm x}_0). 
\end{align}

\end{document}